\documentclass[11pt,a4paper,onecolumn]{article}

\title{Higgs Physics with a
Universal Higgs-Fermion Coupling}
\author{C.K. Bowdery\footnote{E-mail: chris.bowdery@physics.org}
\\Department of Physics, Lancaster University\\Lancaster,
LA1 4YB, UK}
\date{20 December 2002}
\begin{document}

\maketitle
\begin{abstract} A variant of the conventional Higgs model is proposed
which separates the physics of the
Higgs decay modes from the problem of fermion mass generation.
The lowest mass Higgs boson has no significant $b\bar{b}$ decay mode
but a $c\bar{c}$ and $u\bar{u}$  branching ratio of about 33\% each
and requires the existence of two Higgs doublets of type II,
which is consistent with supersymmetric models. It also implies there
are probably no photon or gluon
decay modes of any Higgs boson.
\end{abstract}
\section{General Introduction}

The Standard Model incorporates mass in two different but related ways.
Firstly the $W$ and $Z$ bosons obtain masses by the
spontaneous breaking \cite{higgs1,higgs2,higgs3,higgs4}
of the SU(2)$\times$U(1) electroweak symmetry. Three of the four components
of the complex scalar doublet Higgs field become associated with
the weak bosons while the fourth remains as a scalar particle, the Higgs boson.
The values of the masses are fixed by electroweak couplings.
Secondly the fundamental fermions, quarks and leptons,
obtain their masses by Yukawa couplings \cite{yukawa} to the Higgs field.
Each fermion has its own unique Yukawa coupling constant,
which is not related to the electroweak couplings in any known way.
This means that the values of the fermion masses are
unknown and not computable in the Standard Model. Although
this is an unsatisfactory situation, it is accepted in the absence of any better idea.

The Minimal Supersymmetric Standard Model (MSSM) \cite{MSSM}
has basically the same two mechanisms.
However it incorporates two complex Higgs doublets---one gives
masses to the weak isospin +1/2 fermions ($u$, $c$, $t$ quarks and
possibly the neutrinos) while the other gives masses to the
weak isospin $-1/2$ fermions ($d$, $s$, $b$ quarks and the charged leptons).
In the MSSM there are five unused components of the Higgs fields,
after giving masses to the $W$ and $Z$ bosons,
which leaves three neutral Higgs particles and a pair of charged Higgs particles.

The SM and MSSM do not explain the family (generation)
structure of the fermion masses.
Each fermion has an arbitrary Yukawa coupling, apparently independent
of the family it belongs to and whether it is a quark or lepton.
New physics is required to explain the family structure and the
quark CKM matrix \cite{CKM1,CKM2},
even in the Standard Model.

This note proposes a modification to the fermion mass generation mechanism
which also requires new physics but without specifying what that physics is.
Nevertheless it makes concrete predictions for the decay modes of
the Higgs particles, some of which can be tested with the current data.
The basic philosophy is to separate an intractable problem
into an easy part plus an unknown part in the hope that
an experimental verification of the easy part will lead others
to solve the unknown part.

\section{The Introduction to this Proposal}

We start with the observation that a Higgs-fermion Yukawa coupling
(spin 0 to spin 1/2) is a reasonable thing to have but the existence of 9 (or 12)
different couplings, simply to fit 9 (or 12) different fermion masses, is not.
It would be far simpler and more in the spirit of the rest of particle physics to have
just one, universal, Yukawa coupling. At a stroke this would nearly halve the number
of undetermined constants in the Standard Model.
Of course, this would result in all fermions having the
same mass if the Higgs mechanism were the only contribution to the fermion mass.
By assuming that there are one or more additional contributions to
the fermion mass generation, we can separate the physics of the
branching ratios of the Higgs decay modes from the understanding
of the fermion mass values. Multiple contributions to the observed mass
is already an accepted concept for the proton and neutron and also for electrons
in condensed matter.

A second observation is that the LEP~I and SLC data \cite{indirect} suggest
that a Higgs boson exists with a mass of about 80--90\,GeV/c$^2$,
albeit with large uncertainties.
This is partially in conflict with a direct search for the Higgs particle,
where a limit of about 114\,GeV/c$^2$ has been obtained \cite{LEPlimit} at LEP~II.
However on closer examination, it should be noted that the indirect search result
only really depends on the $Z$-Higgs couplings of the theory while
the direct search result assumes that the Higgs boson
decays predominantly to $b$ quarks.

The proposal in this note does not alter the $Z$-Higgs coupling
so the indirect search result is taken as valid.
However the direct search result is no longer valid if there is a
universal Higgs-fermion coupling constant.

\section{A Universal Higgs-Fermion Coupling in the Standard Model}

With a universal Higgs-fermion coupling, the following decay modes
exist for a low mass Higgs:

\vspace{0.3cm}
\noindent
$\;H^0$ $\rightarrow$  e$^+$e$^-$\hspace{0.5cm}
      $\mu^+\mu^-$\hspace{0.5cm}	$\tau^+\tau^-$\hspace{0.5cm}
      $\nu_e\bar{\nu_e}$\hspace{0.5cm}
      $\nu_{\mu}\bar{\nu_{\mu}}$\hspace{0.5cm}
	$\nu_{\tau}\bar{\nu_{\tau}}$\hspace{0.5cm}
      $u\bar{u}$\hspace{0.5cm}      $d\bar{d}$\hspace{0.5cm}
      $s\bar{s}$\hspace{0.5cm}
      $c\bar{c}$\hspace{0.5cm}	$b\bar{b}$\hspace{0.5cm}
      $t\bar{t}$.

\vspace{0.3cm}
\noindent In order to compute the branching ratios,
we need to give each quark decay mode a weight of three to account for the three
colours of quarks and anti-quarks.
In addition each decay mode should also be doubled for left- and
right-handed spin states.  This then raises the question of whether
this doubling is valid also for the neutrinos.

A few years ago it would have been reasonable to assume that no
neutrino decay modes should exist at all.
However the existence of neutrino oscillations \cite{osc} and the proposed
see-saw mechanism \cite{seesaw1,seesaw2} of neutrino masses has led
to the suggestion
that left-handed neutrinos (and right-handed anti-neutrinos)
might have masses similar to their charged partners and that right-handed neutrinos
(and left-handed anti-neutrinos) might have enormous masses.
So it seems reasonable, now, in a model with a universal
Higgs-fermion coupling to include neutrino decay modes.
The question is whether to incorporate one spin state or two.
This is left for experiments to decide.

For an 80\,GeV/c$^2$ Higgs boson the top quark decay mode is not possible.
This leaves 39 or 42 (approximately) equal decay modes
(counting spins and colour states) with each one then having a
branching ratio of 2.6\% or 2.4\%.
For example:  	

\vspace{0.3cm}
\noindent $\;$BR($H^0 \rightarrow$ any charged lepton pair) = 5\%,
		
\noindent $\;$BR($H^0 \rightarrow$  any $q\bar{q}$) = 14.5\%

\noindent $\;$BR($H^0 \rightarrow$  all $q\bar{q}$) = 72\%

\noindent $\;$BR($H^0 \rightarrow$  all $\nu\bar{\nu}$) = 7.2\% or 14.4\% 

\vspace{0.3cm}
\noindent The fatal flaw with this scheme is the 5\% decay mode to $e^+e^-$.
This would imply that electron positron collisions would produce a
Higgs boson resonance around a centre-of-mass energy of
80\,GeV/c$^2$, which LEP and SLC experiments
have definitely not seen.

However there are a few ways to reconcile a
universal Higgs-fermion coupling with the Standard Model.
One is to assume that the Higgs mass is greater than 200\,GeV/c$^2$
in order to avoid conflicts with unseen $e^+e^-$ resonances but this
is clearly incompatible with the low mass suggested by the indirect
Higgs search data.
Another possibility arises when it is remembered that the Higgs boson has $W$
and $Z$ decay modes as well as fermion/anti-fermion decay modes. In order to
compute the correct branching ratios, it is necessary to know the `baseline mass'
that the universal Higgs-fermion coupling constant provides. 
(This is the degenerate mass all fermions have before additional interactions break
the degeneracy.)

If the `baseline mass' is very large ($> M_Z$), then the fermionic decay modes
will dominate over the weak boson decay modes, whatever the Higgs mass. For
a `baseline mass' about equal to $M_Z$, the weak boson decay modes will
take an equal share (per spin state) with each fermion decay mode
for a Higgs mass of more than 100\,GeV/c$^2$, since off-mass-shell weak boson pair
decays
start to become significant. Still this does not seriously reduce the $e^+e^-$
decay mode. Only if the `baseline mass' is a lot less than $M_Z$ would the weak boson
decay modes dominate and squeeze out the fermionic decay modes.

Although the `very small baseline mass' scenario cannot be ruled out a priori,
it does seem rather contrived and unconvincing when we might expect that the `baseline
mass' should either be close to the weak boson masses, the top quark mass or perhaps
the Higgs energy scale of about 250\,GeV.
Things are quite different for physics beyond today's Standard Model.

\section{The Universal Higgs-Fermion Coupling Concept in
Two Higgs Doublet Models (2HDM)}

In models with two Higgs doublets (2HDM) there are three neutral
Higgs bosons and two charged Higgs bosons.
These are usually labeled $h^0$, $H^0$, $A^0$, $H^-$ and $H^+$.
However we will label the first two neutral Higgs particles
as $H_u^0$ and $H_d^0$ to signify the former couples to the upper-level
(weak isospin +1/2) fermions while the latter couples to the lower-level
(weak isospin $-$1/2) fermions. 
This is the case for type~II 2HDM, of which the MSSM is an example.

To begin with we will ignore any mixing between
these two neutral Higgs bosons, that is, the mixing angle $\alpha=0$.
The fermionic decay modes are:

\vspace{0.3cm}\noindent $\;H_u^0 \rightarrow$  
$\nu_e\nu_e$\hspace{0.5cm}
$\nu_{\mu}\nu_{\mu}$\hspace{0.5cm}
$\nu_{\tau}\nu_{\tau}$\hspace{0.5cm}
$u\bar{u}$\hspace{0.5cm}
$c\bar{c}$\hspace{0.5cm}
$t\bar{t}$\hspace{1cm}and

\noindent $\;H_d^0 \rightarrow$
  e$^+$e$^-$\hspace{0.3cm}
$\mu^+\mu^-$\hspace{0.3cm}
$\tau^+\tau^-$\hspace{0.3cm}
$d\bar{d}$\hspace{0.5cm}
$s\bar{s}$\hspace{0.5cm}
$b\bar{b}$
 
\vspace{0.3cm}
\noindent As with the Standard Model, each listed quark decay mode comes in
three colour states and two spin states;
each lepton decay mode comes in two spin states (or maybe only one each
for the neutrinos).

In order to be compatible with the indirect search result of around 80\,GeV/c$^2$,
the lowest mass Higgs boson must then be $H_u^0$ to avoid having $e^+e^-$ decay modes.
This is the boson that is usually labeled as $h^0$ and which implies
that $\tan\beta$ (the usual ratio of the Higgs vacuum expectation values)
is less than one. There are a number of experimental consequences.

Firstly, for a universal coupling constant,
we would expect an 80\,GeV/c$^2$ $H_u^0$ ($h^0$) to have 15 or 18 decay modes,
counting spin and colour states separately. Thus

\vspace{0.3cm}
\noindent $\;$BR($H_u^0$ $\rightarrow$ $u\bar{u}$)
  = BR($H_u^0$ $\rightarrow$ $c\bar{c}$) = 40\% or 33\%

\vspace{0.3cm}
\noindent There would thus be a branching ratio of 80\% (or maybe 66\%) to hadrons
and a branching ratio of 20\% (or 33\%) to invisible states.
Since $u\bar{u}$ decay modes would be difficult to tag at LEP, only the
charm decay modes would be identifiable and then not very easily.
They could well have gone unnoticed, especially for a Higgs mass close
to that of the $Z$ where the dominant background is $ZZ$.

The second experimental signature would be the clear resonance in $e^+e^-$
collisions at the mass of the $H_d^0$ ($H^0$),
which has to be greater than 200\,GeV/c$^2$,
otherwise it would have been seen at LEP~II.
The $H_d^0$ has 24 decay modes counting spin and colour states separately.
This implies:

\vspace{0.3cm}
\noindent $\;$BR($H_d^0$ $\rightarrow$ $d\bar{d}$) 
  \ \ \ \ = BR($H_d^0$ $\rightarrow$ $s\bar{s}$) 
  \ \ \ \ = BR($H_d^0$ $\rightarrow$ $b\bar{b}$) \ \ \ \ = 25\%\hspace{1cm}and

\noindent $\;$BR($H_d^0$ $\rightarrow$ $e^+e^-$) 
  = BR($H_d^0$ $\rightarrow$ $\mu^+\mu^-$) 
  = BR($H_d^0$ $\rightarrow$ $\tau^+\tau^-$) = 8.3\%.

\vspace{0.3cm}
\noindent The $H_d^0$ would also appear as a resonance if a muon collider were ever
built and in proton-proton and proton anti-proton collisions because
of the $Hd\bar{d}$ coupling.
For hadron colliders the charged lepton decay modes of $H_d^0$
would be obvious signatures.

These straightforward predictions are again subject to modifications depending
on the `baseline mass' of the Higgs mechanism but only if the Higgs mass is more
than about 100\,GeV/c$^2$. For a `baseline mass' of about the $Z$ mass,
the above fermionic decay modes would be reduced as there would be $W/Z$ decay
modes of equal strength. If the `baseline mass' were significantly less than the $Z$
mass then the fermionic decay modes would be replaced by $W$ and $Z$ decay modes.

If there is mixing between $H_u^0$ and $H_d^0$ then the unwanted $e^+e^-$
decay modes of the lightest neutral Higgs boson could again wreck this model.
Although this could be suppressed by choosing different universal Yukawa couplings
for each Higgs doublet---and thus different `baseline masses'---it would seem that
demanding very small or zero Higgs doublet mixing is the only realistic way of
reconciling the universal Higgs-fermion coupling idea with an 80\,GeV/c$^2$ Higgs boson.

The $A^0$ Higgs boson has no weak boson decay modes in a 2HDM of type~II. They decay
to upper-level (weak isospin $+1/2$) fermions proportional to 1/$\tan\beta$
and to lower-level (weak isospin $-1/2$) fermions proportional to $\tan\beta$.
Since $\tan\beta$ is less than one here, this
would imply that the $A^0$  has decay modes similar (but not identical) to the
$H_u^0$
although it might be able to decay to top quarks if its mass is great enough.
Of course, the top decay mode would not dominate as it does in
conventional Higgs models.

The charged Higgs bosons would also have universal couplings to the
fermions giving rise to decay modes very much like $W$ bosons.

\section{Theoretical Implications}

The existence of a Higgs boson with a universal Higgs-fermion coupling
would seem to imply
the need for a two Higgs doublet model (2HDM) of type~II. This could be realised
within a supersymmetric theory and its discovery could provide the first evidence for
SUSY before any SUSY partner particles are observed.

The explanation for the different fermion masses---the breaking of the mass
degeneracy provided by this variant of the Higgs mechanism---clearly
requires new physics.
(Of course, unless the Yukawa coupling constants are taken as inexplicable
fundamental constants, a similar statement is also true in
the conventional SM and MSSM.) This paper does not attempt to propose what
the new physics would be. However it is possible to outline some of the ingredients and
consequences of the new physics. The following discussion assumes a SUSY theory.

Firstly, it seems reasonable to assume that both fermions and their SUSY partners
have the same universal Higgs coupling and thus all have the same `baseline mass',
perhaps around several hundred GeV/c$^2$. Some new interaction would then act upon
the R-parity quantum number and
lower the `normal' fermions to a smaller mass and lift the
SUSY partner particles to a larger mass. A second stage would presumably
cause further shifts
(`splitting') according to the family (generation) quantum number, of which little
is currently known. The final particle masses would presumably
appear after additional shifts
caused by colour and electroweak interaction effects.

It would be natural that quarks with their colour interactions would have bigger
masses than leptons in a given family/generation. Furthermore, charged leptons would be
expected to be heavier than neutral leptons because of electric interactions and similarly
charged 2/3 quarks would be expected to be heavier than charge 1/3 quarks.
That this is not true for \textsl{up} and \textsl{down} could be due to
$Z$ boson effects overwhelming photon effects, since the so-called 1st generation quarks
are shifted by a very large energy amount from the `baseline mass' value.

If the fermions and their SUSY partners do share the same universal coupling then,
for Higgs interactions, SUSY is effectively an unbroken symmetry. This implies
that any Feynman diagram with a Higgs boson coupling to a fermion loop will have a
cancelling diagram with a SUSY partner loop. This cancellation will be exact.
The consequence of this is 
the elimination of all higher order loop corrections to the allowed decay modes
and also the removal of all Higgs decay modes
to photons and gluons. (This latter effect has profound implications for experimental
physics searches for the Higgs.)

Having no Higgs decay modes to photons and gluons removes
any possibility for truly massless particles to interact with the
Higgs field in the vacuum. This is clearly self-consistent. (Could it be
that the Standard Model,
and conventional broken SUSY, without this exact cancellation of loops, is not
self-consistent?) 

\section{Alternative Variations}

This proposal has dealt with a single, universal, Yukawa coupling constant scenario.
The conventional view is that there are 9 or 12 such constants.
Clearly there is room for any
number of Yukawa constants from zero up to 9 or 12.

Zero constants might seem an attractive idea with the Higgs mechanism having no
direct part in generating fermion masses. In the Standard Model there might still
be an indirect effect via weak boson loops although there might be cancellations
in a SUSY model. However the main reason for not developing this
idea in this paper is the reasonableness of having a Yukawa coupling.
Some explanation for removing it would need to be found.

Two Yukawa couplings is still a possibility in a 2HDM---one constant for
each Higgs doublet. Three Yukawa couplings might make sense with one for
each family/generation. However it would probably imply a large $b$ quark decay
mode, incompatible with an 80\,GeV/c$^2$ Higgs boson. Four Yukawa
constants could give masses for each fermion in a family with
other interactions splitting each family. However this idea probably founders
because there is no clear repeated pattern of masses seen within the three families.

\section{Conclusions}

The Standard Model and conventional SUSY variants of the Standard Model
incorporate multiple Yukawa couplings between Higgs bosons and fermions
in order to account for the different fermion masses.
This paper proposes that there is only a single, universal Higgs-fermion
coupling constant. In order to be compatible with experimental data, this concept
appears to require at least a two Higgs doublet model of type~II, such as a
supersymmetric theory.

If so, the lightest Higgs boson, $h^0$, could be (but does not have to be) lighter than
the currently accepted limits, possibly as low as 80\,GeV/c$^2$.
Its visible decay modes would be to
\textsl{up} and \textsl{charm}, each of about 33\%, or possibly 40\%
if there are no couplings to
right-handed neutrinos. The heavier neutral Higgs boson, $H^0$,
should be easy to detect in electron positron,
muon and hadron colliders and its mass is greater than 200\,GeV/c$^2$.
The charged lepton pair decay modes with branching ratios of 8.3\%
should be straightforward, whatever the collider.
The $b\bar{b}$ decay mode at 25\% could be exploited using $b$-tagging techniques.
There would probably be no decay modes to photons or gluons.

In a sense, this paper has advocated \emph{less} physics, namely only a single
Yukawa coupling constant. However it does imply the existence of new physics
to explain the breaking of the fermion mass degeneracy. Since even the Standard
Model requires new physics to explain the values of the multiple Yukawa constants,
that fact ought not to count against this proposal.
\section*{Acknowledgements}
I would like to thank the following for assistance and encouragement: 
Kate Fenton, Tony Gu\'enault, Rich Haley, Stephen Haywood, Holly Thomas and Anne Watson.


\begin{thebibliography}{99}
\bibitem{higgs1}P.~Higgs, Phys. Lett. \textbf{12} (1964) 132,
Phys. Rev. Lett. \textbf{13} (1964) 508,
Phys. Rev. \textbf{145} (1966) 1156
\bibitem{higgs2}F.~Englert \& R.~Brout, Phys. Rev. Lett. \textbf{13} (1964) 321
\bibitem{higgs3}
G.S.~Guralnik, C.R.~Hagen \& T.W.B.~Kibble, Phys. Rev. Lett. \textbf{13} (1964) 585
\bibitem{higgs4}
T.W.B.~Kibble, Phys. Rev. \textbf{155} (1967) 1554
\bibitem{yukawa}J.F.~Gunion et al.,\textsl{The Higgs Hunter's Guide}
Addison-Wesley (1990)
\bibitem{MSSM}P.~Fayet Nucl Phys \textbf{B90} (1975) 104
\bibitem{CKM1}N.~Cabibbo, Phys. Rev. Lett. \textbf{10} (1963) 531
\bibitem{CKM2}
M.~Kobayashi \& K.~Maskawa, Progr. Theor. Phys \textbf{49} (1972) 282
\bibitem{indirect}CERN-EP-2000-016, LEP Collaborations; updated in A.~Straessner's talk at
the 2000 Electroweak Rencontres de Moriond
\bibitem{LEPlimit}CERN-EP/2001-055, LEP Higgs working group
\bibitem{osc}The SNO Collaboration, Phys. Rev. Lett. \textbf{89} No 1 (2002) 011301
\bibitem{seesaw1}M.~Gell-Mann, P.~Ramond \& R.~Slansky, \textsl{Supergravity}
ed D.~Freedman et al., North Holland (1979)
\bibitem{seesaw2}T.~Yanagida, Prog. Theor. Phys. \textbf{64} (1980) 1103
\end{thebibliography}
\end{document}